\documentclass{emulateapj}

\usepackage{apjfonts}

\def\snr{G350.1--0.3}
\def\xmm{{\em XMM-Newton}}
\def\xm{{\em XMM}}
\def\cxo{{\em Chandra}}
\def\rxj{1RXS~J172106.9--372639}
\def\ax{AX~J1721.0--3726}
\def\cco{XMMU~J172054.5--372652}
\def\kms{km~s$^{-1}$}
\def\la{\ifmmode\stackrel{<}{_{\sim}}\else$\stackrel{<}{_{\sim}}$\fi} 
\def\ga{\ifmmode\stackrel{>}{_{\sim}}\else$\stackrel{>}{_{\sim}}$\fi} 
\newcommand\HI{H\,{\sc i}}

\def\kms{km~s$^{-1}$}
\def\etal{{\rm et~al.\ }}

\shorttitle{THE YOUNG, LUMINOUS SUPERNOVA REMNANT \snr}
\shortauthors{GAENSLER ET AL.}

\begin{document}

\title{The (Re-)Discovery of G350.1--0.3: A Young, Luminous Supernova Remnant and Its Neutron Star}
\submitted{Accepted to {\em The Astrophysical Journal (Letters)}}
\author{B. M. Gaensler,\altaffilmark{1,8} A. Tanna,\altaffilmark{1},
P. O. Slane,\altaffilmark{2} C. L. Brogan,\altaffilmark{3}
J.~D.~Gelfand,\altaffilmark{4,9} \\
N. M. McClure-Griffiths,\altaffilmark{5} 
F.~Camilo,\altaffilmark{6}
C.-Y. Ng,\altaffilmark{1} and
J. M. Miller\altaffilmark{7}}
\altaffiltext{1}{School of Physics, The University
of Sydney, NSW 2006, Australia}
\altaffiltext{2}{Harvard-Smithsonian
Center for Astrophysics, Cambridge, MA 02138}
\altaffiltext{3}{National Radio Astronomy Observatory,
Charlottesville, VA 22903}
\altaffiltext{4}{Center for Cosmology and Particle Physics,
New York University, New York, NY 10023}
\altaffiltext{5}{CSIRO Australia Telescope National Facility,
Marsfield, NSW 2122, Australia}
\altaffiltext{6}{Columbia Astrophysics Laboratory, Columbia University,
New York, NY 10027}
\altaffiltext{7}{Department of Astronomy, University of Michigan, 
Ann Arbor, MI 48103}
\altaffiltext{8}{Australian Research Council Federation Fellow}
\altaffiltext{9}{Astronomy and Astrophysics Fellow, National Science
Foundation}

\begin{abstract} 
We present an \xmm\ observation of the long-overlooked radio source
\mbox{\snr}.
The X-ray spectrum of \snr\ can be fit by a
shocked plasma with two components: a high-temperature (1.5~keV)
region with a low ionization time scale and enhanced abundances,
plus a cooler (0.36~keV) component in ionization
equilibrium and with solar abundances. The X-ray spectrum and the
presence of non-thermal, polarized, radio emission together demonstrate that
\snr\ is a young, luminous supernova remnant (SNR),
for which archival \HI\ and $^{12}$CO data indicate
a distance of 4.5~kpc.  The diameter of the source then implies an age of
only $\approx 900$~years. The SNR's distorted appearance,  small
size and the presence of  $^{12}$CO emission along the SNR's eastern
edge all indicate that the source is interacting with a complicated
distribution of dense ambient material.  An unresolved X-ray source,
\cco, is detected a few arcminutes west of the brightest SNR emission.
The thermal X-ray spectrum and lack of any multi-wavelength counterpart
suggest that this source is a neutron star associated with \snr,
most likely a ``central compact object'', as
seen coincident with other young SNRs such as Cassiopeia~A.
\end{abstract}


\section{Introduction}

Many supernova remnants (SNRs) expand into complex environments
shaped by molecular clouds and by the strong winds from OB associations.
If our understanding of SNRs is to progress, we must 
aim to identify and understand these complicated systems.

\snr\ is a bright radio source in the inner Galaxy.
Its linear polarization and non-thermal spectrum 
originally led to its classification
as a SNR \citep{ccg73,ccg75}, but a high-resolution image \citep{spsh86}
revealed a distorted, elongated, morphology, very
different from the shell structure usually seen in such sources.
\cite{spsh86} argued that \snr\ was possibly a radio galaxy or a
galaxy cluster.  In subsequent SNR catalogs, the source was either
downgraded to a SNR candidate or dropped completely \citep{gre91,wg96},
and has subsequently been forgotten.
We here present new and archival data on this unusual source. We
demonstrate that \snr\ is a very young and luminous SNR expanding
into dense ambient gas, and reveal an associated neutron star likely
to be an addition to the growing class of ``central compact objects''
(CCOs).

\section{Observations}

\begin{figure}
\centerline{
\includegraphics[width=0.6\textwidth]{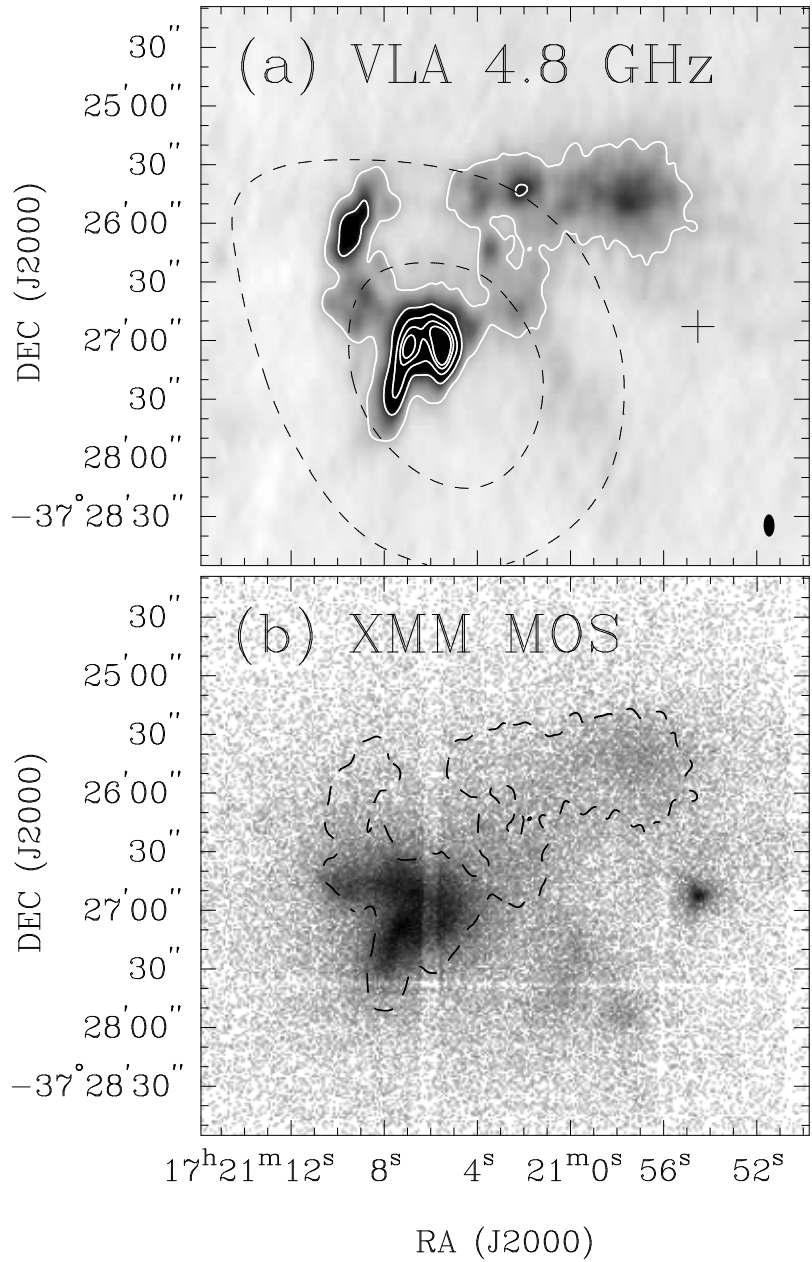}}
\caption{Radio and X-ray observations of \snr. Panel (a) shows a
4.8~GHz image at a resolution of $11\farcs4\times 5\farcs6$
(indicated by the ellipse at lower right) and a sensitivity of
170~$\mu$Jy~beam$^{-1}$, generated from archival VLA data.
The greyscale is linear between --1 and +10~mJy~beam$^{-1}$; white
contours are at levels of 2, 8, 16, 24 \& 30~mJy~beam$^{-1}$.
The dashed contours show {\em ROSAT}\ All-Sky Survey data smoothed
to $2'$ and with contours at 50\% \& 80\% of the peak. The cross
marks the position of \cco. Panel (b) corresponds to the sum of
\xm\ EPIC MOS1 and MOS2 data in the energy range 0.5--10~keV, with
a logarithmic greyscale in units of counts. The dashed contour
corresponds to the VLA data from panel (a) at the level of
2~mJy~beam$^{-1}$.}
\label{fig_image}
\end{figure}

In Figure~\ref{fig_image}(a) we show a 4.8~GHz radio image of \snr,
generated from Very Large Array (VLA) data taken 
in 1984 Jun 06 (C configuration; project code AV105),
1984 Aug 27 (D configuration; AV105) and 1989 Jun 23 (CnB configuration; AB544).
The source
is dominated by a bright double-peaked clump,
with diffuse, elongated, structures extending 
to the north-east and north-west of this.
\snr\ has been detected in X-ray surveys with
{\em ROSAT}\ and {\em ASCA},
in which it was designated \rxj\ and \ax, respectively
\citep{vab+99,smk+01}.  These data provide limited spatial information
(see Fig.~\ref{fig_image}[a]), but the $\sim1000$ counts from {\em
ASCA}\ are a good fit to a non-equilibrium ionization (NEI) plane-parallel
shocked plasma
with variable abundances \citep[XSPEC model 
``VPSHOCK'';][]{blr01}, with absorbing column $N_H \approx
3\times10^{22}$~cm$^{-2}$, temperature $kT \approx 1.6$~keV, an
ionization time scale $\tau \approx 3\times10^{11}$~s~cm$^{-3}$,
an absorbed 0.5--10~keV flux
$\approx10^{-11}$~ergs~cm$^{-2}$~s$^{-1}$, and enhanced
levels of Si, S and Fe.

Data from the Southern Galactic Plane Survey \citep{mdg+05} show
that there is strong \HI\ absorption seen against \snr\ at velocities
in the range 0 to --40~\kms\ relative to the local standard of rest
(LSR). However, no absorption is seen at --80~\kms, despite the
presence of bright ($T \ga 30$ K) \HI\ emission at this velocity
and in this direction. The systemic velocity for \snr\ thus must
fall in the range  --80 to
--40~\kms, ruling out an extragalactic origin
and implying a distance\footnote{We assume standard IAU
parameters for the solar orbital velocity, $\Theta_0 = 220$~\kms,
and for the distance to the Galactic Center, $R_0 = 8.5$~kpc.}
between 4.5 and 10.7~kpc.
In further discussion we adopt a distance to \snr\ of
$4.5d_{4.5}$~kpc; in \S\ref{sec_snr} we will argue that $d_{4.5}
\approx 1$.

We observed \snr\ with \xmm\ for 35~ks on 2007 Feb 23.  The EPIC
pn CCD was operated in full-frame mode, while EPIC MOS1 and MOS2
were used in large-window mode.  The data were screened to remove
hot pixels and flares, resulting in effective exposures of 34.7~ks
(MOS) and 29.6~ks (pn).

\section{Results}

An \xm\ image of \snr\ is shown in Figure~\ref{fig_image}(b).
Overall, the source is extremely bright: the background corrected
count rates (0.5--10~keV) are $1.436\pm0.007$~counts~s$^{-1}$ (MOS1),
$1.309\pm0.007$~counts~s$^{-1}$ (MOS2) and $2.97\pm0.01$~counts~s$^{-1}$
(pn), corresponding to $>180\,000$ events from all three cameras
combined.  The source is dominated by a bright X-ray clump in
the south-east, coincident with the region of brightest radio emission.

A notable difference between the two wavebands is the bright
unresolved X-ray source $\approx140''$ west of the brightest radio
and X-ray emission. This source is at RA $17^{\rm h}20^{\rm m}54\fs5$,
Decl.\ $-37^\circ26'52''$ (with an approximate uncertainty of
$\pm2''$ in each coordinate), and we correspondingly designate it
\cco.  There is no radio counterpart in the VLA image, nor is there
any optical or infrared source at this position in the Digitized
Sky Survey, 2MASS or GLIMPSE.  However, the optical and infrared
limits are unconstraining given the heavy confusion and high
extinction, along with the relatively large uncertainty of the \xm\
position.

\subsection{Diffuse Emission}
\label{sec_diffuse}

\begin{figure}
\centerline{
\includegraphics[width=0.8\textwidth]{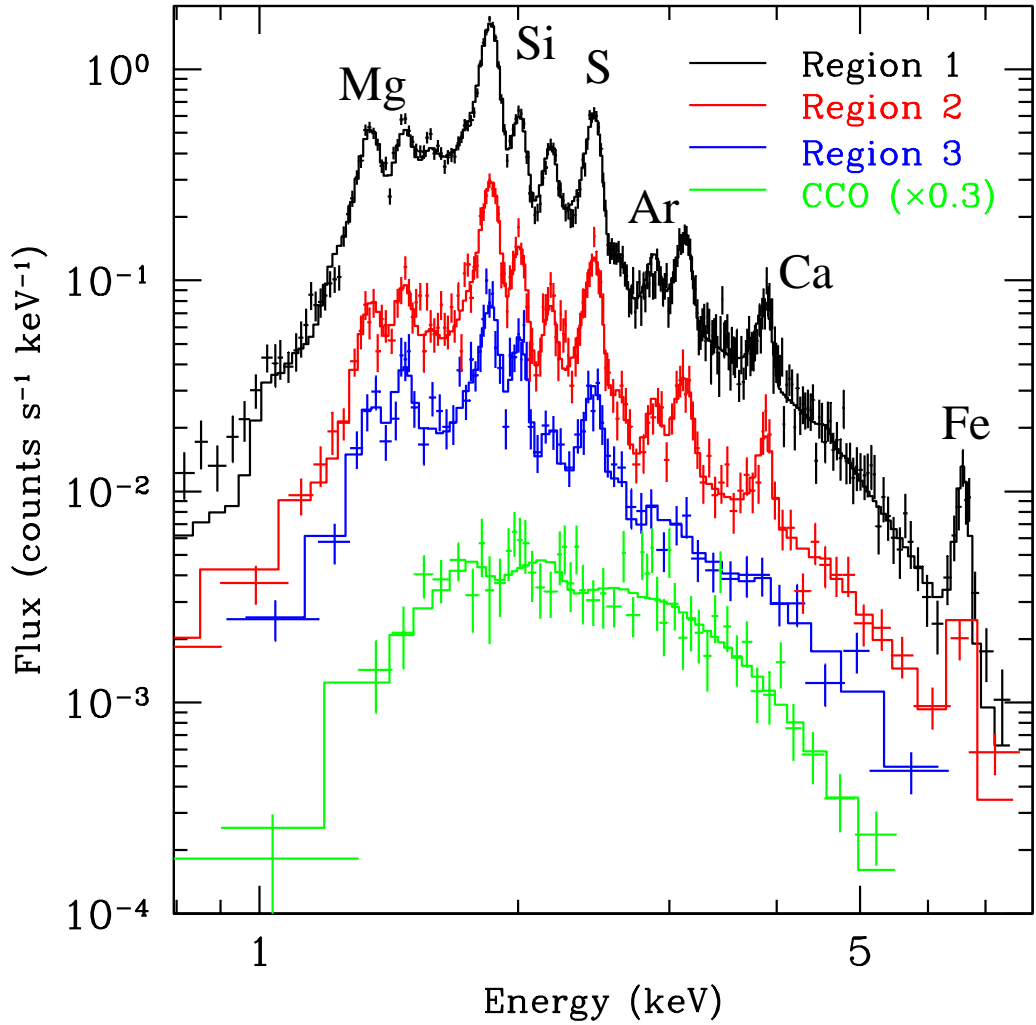}}
\caption{\xm\ spectra of \snr. The black spectrum
corresponds to EPIC pn data from the brightest diffuse region, extracted
from a circle of radius $48''$, centered on (J2000) RA $17^{\rm
h}21^{\rm m}07\fs44$, Decl.\ $-37^\circ27'01\farcs8$; several atomic
emission lines are apparent, as indicated. The red and blue dara
show EPIC~MOS1 spectra
from two adjacent sub-regions within the region shown in black (see text
for details). The green spectrum
is that of EPIC~pn data within a $21''$-radius circle centered on the unresolved
source \cco.  In all cases, the points show
the data after appropriate background subtraction, while the solid
lines represent the corresponding best-fit models as discussed in the
text.}
\label{fig_spec}
\end{figure}

The X-ray spectrum of the brightest region, shown in
Figure~\ref{fig_spec}, is rich with line emission. No single thermal
model can be sensibly fit to these data, but they can be adequately
described by a two-component model, consisting of a low-temperature
collisionally equilibrated plasma with solar abundances \citep[XSPEC
model ``RAYMOND'';][]{rs77}, plus a higher-temperature component with
a low ionization time scale (``VPSHOCK'') and with large over-abundances of all
metals beyond neon (below which constraints cannot be obtained due
to the large column density).  The corresponding best joint fit to
MOS1, MOS2 and pn data is shown in black in Figure~\ref{fig_spec} (only the pn
spectrum is plotted), with parameters as listed in Table~\ref{tab_spec}.
Most of the significant residuals are below 1~keV, and indicate a problem
with the model in the Fe-L region of the spectrum. An ad-hoc NEI model
consisting only of an iron line (plus continuum) can further improve
the fit, possibly suggesting a range of ionization time scales for Fe
throughout the region.  Spectra of the 
diffuse emission extending throughout the source are similar to
that seen from the brightest region.

Even at the limited angular resolution of \xm, there is spectral
structure within the brightest X-ray-emitting region. Extraction
of separate sub-regions within this feature in circles of radius
$12''$ reveal spectra that are each well-fit by a single absorbed
VPSHOCK model, with temperatures and column densities consistent
with those derived in Table~\ref{tab_spec} for the entire bright
region. However, there are vast differences between regions in both
abundances and ionization states. For example, a MOS1 spectrum from
a $12''$-radius circle centered
at RA $17^{\rm h}21^{\rm m}07$, Decl.\ $-37^\circ26'46''$ (shown
in red in Fig.~\ref{fig_spec}) shows higher abundances in Si, Ar
and Ca than emission $30''$ to the east of this (shown in blue
in Fig~.\ref{fig_spec}).  Comparison of the He-like and H-like lines
of Si indicates significant differences between the ionization
levels of these two regions also.

\subsection{The Compact Source}

In contrast to the surrounding diffuse emission, the \xm\ spectrum
of \cco\ shows a featureless continuum, as  shown in green in 
Figure~\ref{fig_spec}.
These data cannot be described by the models considered in
\S\ref{sec_diffuse}. An absorbed power law yields an acceptable fit
($\chi^2$/dof~$=1.16$) but with an unphysically steep photon index,
$\Gamma = 5.4$.  The best simple fit ($\chi^2$/dof~$=1.06$) is
provided by an absorbed blackbody, with $N_H =
(2.9\pm0.3)\times10^{22}$~cm$^{-2}$, $kT = 0.53\pm0.02$~keV and an
unabsorbed 0.5--10~keV flux
$(1.4_{-0.4}^{+0.5})\times10^{-12}$~ergs~cm$^{-2}$~s$^{-1}$ (errors
all at 90\% confidence). 

The MOS data have a time-resolution of 0.9~s, while the pn events
have a resolution of 73.4~ms. Within these constraints, we can
search for X-ray pulsations from \cco. We extracted 1378 MOS events
and 922 pn events within $7\farcs5$  of the source and within the energy
range 1--5~keV.  The arrival times were shifted to the solar system
barycenter, and a pulsation search was carried out using the $Z_n^2$
test \citep{bbb+83}.  No significant periodic signals were found
for $n=1,2,3$.  Assuming a sinusoidal pulse, the corresponding
99.999\% confidence upper limits on the 1--5~keV pulsed fraction
are $\sim33\%$ for periods in the range 146~ms to 1.8~s,
and $\sim20\%$ for periods from 1.8~s up to 1~hour.

\section{Discussion}

\subsection{\snr: A Young, Luminous SNR}
\label{sec_snr}

\snr\ is extended, Galactic, and emits both polarized,
non-thermal, radio emission and thermal X-rays with enhanced
abundances. Despite its unusual morphology, its emission properties
unquestionably demonstrate that it is a supernova remnant.  The
source diameter $D\approx 2.6d_{4.5}$~pc and unabsorbed 0.5-10~keV
luminosity $L_X \approx 2d_{4.5}^2\times 10^{36}$~ergs~s$^{-1}$ put
this long over-looked source among the smallest and most luminous
few SNRs in the Galaxy.

As is the case for other young SNRs \citep[e.g.,][]{lsh+05,msb+07},
the high- and low-temperature components of the X-ray spectrum can
be broadly interpreted as shocked ejecta and swept-up ambient
material, respectively. The swept-up component has reached collisional
equilibrium while the ejecta are still ionizing. The temperature
of the ambient gas implies a shock velocity $V_s = 560^{+20}_{-40}$~\kms,
assuming equilibration between ions and electrons.  If we assume
that the SNR is in the Sedov phase, we can estimate an age\footnote{If
electron-ion equilibration has not yet been achieved, the true age
is even lower than this estimate.} age $t_{SNR} = D/5V_s \approx
900d_{4.5}$~years, with the caveat that an accurate diameter is
difficult to extract from the complex morphology.

Two independent calculations imply a high ambient density into which
this SNR is expanding.  First, the fact that the cooler shock
component has reached collisional ionization equilibrium implies
an ionization time scale $\tau \equiv n_e t_{SNR} \ga 3
\times10^{12}$~s~cm$^{-3}$ \citep{mas94,blr01}, where $n_e$ is the
electron density of the swept-up material. This lower limit then
implies an ambient density $n_0 \approx n_e/4 \ga 25d_{4.5}^{-1}$~cm$^{-3}$.
Second, standard Sedov expansion for $D = 2.6d_{4.5}$~pc and $t_{SNR}
\approx 900d_{4.5}$~years implies $n_0/E_{51} \approx
600d_{4.5}^{-3}$~cm$^{-3}$,
where $E_{51}\times10^{51}$~ergs is the kinetic energy of the
supernova explosion. For $E_{51} \approx 1$,
the ambient density is again very high.

These results suggest that the SNR is interacting with dense,
molecular, gas.  The strange source morphology provides
additional evidence for a molecular cloud interaction, since such
encounters can produce SNRs with asymmetric 
and highly distorted appearances
\citep[e.g.,][]{tby85,wrm98}.  In this case, the dense
material is expected to  sit east of the SNR, adjacent to
the brightest radio/X-ray emission.  Support
for this conclusion is provided by the $^{12}$CO survey of
\cite{bab+97}, which shows that immediately to the east of \snr\
is a clump of molecular gas at LSR velocity --40~\kms.  We
propose that \snr\ is associated with this cloud; the 
distance to the system is then 4.5~kpc, so that $d_{4.5} \approx 1$.

\subsection{\cco: A Candidate CCO}

\cco\ is an unresolved thermal X-ray source with no radio counterpart,
in close proximity to a young SNR. We thus suggest
that this source is a neutron star associated with \snr. This would require
that \snr\ be the result of a core-collapse supernova (as also implied
by the evidence for interaction with a molecular cloud, as discussed
in \S\ref{sec_snr}).
There are a wide variety of categories into which young neutron stars
can fall. The absence of any non-thermal component to the X-ray spectrum
of \cco\ rules out magnetospheric or nebular emission associated with a
young, rotation-powered pulsar.  Furthermore, while young pulsars can also
show  thermal surface emission, the blackbody temperature of these X-rays
is typically $kT \la 0.1$~keV, much lower than seen here \citep{krh06}.

The observed surface temperature $kT \approx 0.5$~keV and 0.5--10~keV
luminosity $L_X \approx 3d_{4.5}^2 \times 10^{33}$~ergs~s$^{-1}$
of \cco\ are both within the range seen for magnetars \citep{wt06}.
However, the X-ray spectra of magnetars show hard power-law
tails, unlike what is seen here.  Magnetars show slow ($\sim$2--12~sec)
X-ray pulsations, often with a significant ($\ga40\%$) pulsed
fraction.  While the lack of slow pulses in our data provides
initial evidence against such an identification, deeper pulsation
searches will be needed to fully constrain such behavior.
A more likely possibility is that \cco\ is a
central compact object \citep[see][for a 
review]{del08}. This group of $\sim 10$ sources are all associated
with young SNRs. They are all thermal X-ray sources, with X-ray
luminosities $\sim10^{33}$~ergs~s$^{-1}$, and blackbody temperatures
$kT\approx0.3-0.5$~keV. Two CCOs are pulsed \citep[at periods of
105 \& 424~ms;][]{gh08}; the rest show no X-ray variability.  No CCOs
show counterparts in other wavebands.

\cco\ appears to meet all these criteria; we thus propose
that it is a CCO associated with \snr. One potential difficulty
with this claim is that within the formal statistical
errors, the absorbing columns of the SNR and of the compact
source are inconsistent. However, there is uncertainty as to the
true spectral shape of both sources: in the case of the SNR because
of unmodeled residuals in the spectrum below 1~keV, and for
\cco\ as a result of likely atmospheric effects on the neutron star
surface (which we have not attempted to fit for here). With these
systematic effects included, the foreground columns of the two
sources are sufficiently close that they can be considered consistent.
In any case, the proposed interaction with molecular material may
lead to large position-dependent fluctuations in absorption.

Other CCOs are found very close to the geometric centers of their
associated SNRs.  In contrast, \cco\ sits at the western extreme
of the extended radio and X-ray emission, suggesting that it
received a substantial kick in the supernova explosion.  
If we identify the centroid of the lowest radio contour in Figure~\ref{fig_image}
as the explosion site, the offset of the CCO from the SNR center is
$\approx2d_{4.5}$~pc, requiring a projected space velocity for the neutron
star of $\approx2000$~\kms\ (independent of $d_{4.5}$).  We consider this
unlikely, since this value is well above the 330~\kms\ measured for the
CCO in Cassiopeia~A \citep{tfv01}, and exceeds even the 1100--1600~\kms\
projected velocity measured for the CCO in Puppis~A \citep{hb06,wp07}.
An alternative is that the CCO is still reasonably close to the site
of the original supernova, and that a considerable portion of the
SNR (farther to the west) remains unseen in both the radio and X-ray
bands. Although there is a proposed molecular cloud interaction to the
east, ejecta traveling westward may be expanding relatively unimpeded into
a low density environment.  Another possibility is that \snr\ is actually
two or more overlapping or interacting SNRs \citep[e.g.,][]{wcd+97},
and that \cco\ is at the center of a partial shell corresponding to the
fainter, western, half of the overall X-ray structure.

\section{Conclusion}

New X-ray data demonstrate that \snr\ is a very young
($\sim900$ year old) and luminous supernova remnant, most likely
interacting with a molecular cloud at a distance of 4.5~kpc.  We
also identify the compact thermal X-ray source \cco, which we propose
as a central compact object  associated with \snr.

The complicated morphology and spectrum of the SNR make it difficult
to carry out detailed calculations of its properties.  For example,
the spectrum of the diffuse emission that extends throughout the
rest of the remnant is consistent with that from the brightest
region, indicating that ejecta fills the X-ray emitting regions.
However, this leaves open the question of why there is such a bright
concentration in one small region of the SNR.  Higher angular
resolution observations of the SNR with the {\em Chandra X-ray
Observatory}, combined with further millimeter observations
of the environment into which the source is expanding, are needed
to better address such issues.  The nature of \cco\ also needs
further confirmation --- deeper \xm\ observations at high time
resolution can provide better constraints on pulsations, while a
subarcsecond localization from \cxo\ can enable a deep search
for an infrared counterpart.

Clearly, \snr\ is one of a growing number of ``missing'' SNRs needed
to balance the estimated supernova rate for the Milky Way of one
per $\sim50$~years. It is noteworthy that
at least four such sources, G1.9+0.3 \citep{rbg+08}, G12.8--0.0
\citep{bgg+05}, G327.2--0.1 \citep{gg07} and now G350.1--0.3, have
all been recently revealed to be very young SNRs as a result of
follow-ups to X-ray or $\gamma$-ray detections.  Further
investigation of unidentified Galactic high-energy sources
will likely add further to this exciting sample
of objects.

\begin{acknowledgements}
We thank John Hughes and Rob Fesen for helpful discussions, and the
Lorentz Centre at Leiden University where some of this work was
carried out.  We acknowledge the support of NASA through grants
NNX06AH60G \& NAG5-13032 (B.M.G.) and contract NAS8-39073
(P.O.S.).  NRAO is a facility of the NSF, operated under cooperative
agreement by AUI.
\end{acknowledgements}

{\it Facilities:} XMM (EPIC), VLA


\begin{deluxetable}{llc}
\tabletypesize{\scriptsize}
\tablewidth{0pt}
\tablecaption{Spectral fit of a two-temperature shocked plasma to
the region of brightest X-ray emission in \snr.\label{tab_spec}}
\tablehead{  & VPSHOCK  & RAYMOND}
\startdata
$N_H$\tablenotemark{a}  &
\multicolumn{2}{c}{$3.71^{+0.05}_{-0.09}\times10^{22}$~cm$^{-2}$} \\
$kT$ (keV) & $1.46^{+0.09}_{-0.06}$ &  $0.36^{+0.02}_{-0.05}$ \\
$\tau$  & $3.0^{+0.3}_{-0.4}\times10^{11}$~s~cm$^{-3}$ & \nodata \\
Ne &  $40^{+70}_{-20}$ & (1) \\
Mg &  $20^{+140}_{-10}$ & (1) \\
Si &  $20^{+90}_{-10}$ & (1) \\
S &  $12^{+70}_{-5}$ & (1) \\
Ca & $18^{+5}_{-7}$ & (1) \\
Fe &  $6^{+7}_{-3}$ & (1) \\
Flux\tablenotemark{b}  &
\multicolumn{2}{c}{$(6-14)\times10^{-10}$~ergs~cm$^{-2}$~s$^{-1}$} \\
$\chi^2$/dof & \multicolumn{2}{c}{1679/1135=1.48}
\enddata

\tablecomments{The three spectra from EPIC MOS1, MOS2 and pn were
fit jointly to an absorbed VPSHOCK + RAYMOND spectral model, except
for the overall normalization, which was fit independently for each
instrument.  Uncertainties are all statistical errors at 90\%
confidence, except for the unabsorbed flux, for which the range
given reflects systematic uncertainties between the fluxes measured
with MOS1, MOS2 and pn.  Abundances are relative to solar; entries
in parentheses indicate quantities held fixed at solar values.}

\tablenotetext{a}{\scriptsize Interstellar absorption has been calculated
using the model of \cite{wam00} assuming solar abundances, and has
been applied simultaneously to both temperature components.}

\tablenotetext{b}{\scriptsize The flux is the sum of both components
over the energy range 0.5--10.0~keV, and has been corrected for
foreground absorption.}

\end{deluxetable}
\normalsize

\clearpage

\end{document}